\documentclass{PoS}

\usepackage{graphicx}
\usepackage{bm}
\usepackage{amsmath}
\usepackage{amssymb}

\title{Charmonium spectra and dispersion relation with improved Bayesian
analysis in lattice QCD}

\ShortTitle{Charmonium dispersion relation with MEM in lattice QCD}

\author{\speaker{Atsuro Ikeda}\\
Department of Physics, Osaka University , Toyonaka, Osaka 560-0043, Japan\\
        E-mail: \email{a-ikeda@kern.phys.sci.osaka-u.ac.jp}}

\author{Masayuki Asakawa\\
Department of Physics, Osaka University , Toyonaka, Osaka 560-0043, Japan\\
        E-mail: 
        \email{yuki@phys.sci.osaka-u.ac.jp}}
\author{Masakiyo Kitazawa\\
Department of Physics, Osaka University , Toyonaka, Osaka 560-0043, Japan\\
        E-mail: 
        \email{kitazawa@phys.sci.osaka-u.ac.jp}}

\abstract{
We study the charmonium spectral functions at finite momentum 
and the dispersion relation of $\eta_c$ at finite temperature.
For the analysis of the spectral function, we use an extended 
maximum entropy method (MEM).
We perform the MEM analysis for the product space 
of Euclidean correlators in different channels or 
momenta to incorporate information encoded in correlations
among the Euclidean correlators in MEM.
We find that this method can improve the error of
the reconstructed spectral images.
To study the dispersion relation, we introduce a definition 
of the peak position in the spectral image in which the  
associated error can be estimated on the basis of MEM.
We find that the dispersion relation of $\eta_c$ at finite temperature
follows the Lorentz invariant form even near 
the dissociation temperature $T\simeq1.7T_c$.
}

\FullConference{The 32nd International Symposium on Lattice Field Theory,\\
		23-28 June, 2014\\
		Columbia University New York, NY}

\begin{document}
\section{\label{sec:intro}Introduction}
The Quark-Gluon Plasma (QGP) near the pseudo critical temperature 
($T_c$) is believed to be a strongly 
interacting liquid-like matter.
Lattice QCD is a powerful method to study 
properties of the QGP in such a non-perturabative environment.
The dynamical properties of QGP near $T_c$ are among the most interesting subjects in this field.
The spectral function is a useful quantity to study the dynamical properties of the system.
A peak in the spectral function represents the existence of a bound
state \cite{asakawa_j/_2004,datta_behavior_2004}.
The vector spectral function is related to  the thermal dilepton production rate
\cite{karsch_lattice_2002, ding_thermal_2011} and its slope at the origin is related to transport
properties according to the Kubo formulae \cite{karsch_thermal_1987}.

On the lattice, however, directly we can only calculate the imaginary time 
correlation functions.
Thus, we must perform the analytic continuation to obtain the spectral 
function, which is the imaginary part of the real time correlator.
The solution to this problem is non-trivial, 
because it is impossible to reconstruct a continuum function from a limited
number of data points without some assumptions \cite{jarrell_bayesian_1996, asakawa_maximum_2001}.
Maximum Entropy Method (MEM) \cite{jaynes_information_1957} is one of the well-known
methods to deal with this problem.
With MEM, spectral functions are reconstructed from the 
lattice correlator and prior knowledge.
One of the advantages of MEM is that this method enables us to estimate the
probabilistic error of the reconstructed image.

The charmonia created in heavy ion collisions move through the QGP medium.
However, only charmonia with zero momentum have been mainly considered on the
lattice in the previous research \cite{aarts_charmonium_2007, ding_charmonium_2012-1}
(for some exceptions, see, for example \cite{ding_momentum_2013}).
The purpose of the present study is to 
investigate dynamical properties of charmonia in the pseudoscalar
channel at finite momenta at finite temperature ($T$).
To this end, we analyze the spectral function $A(\omega,\vec{p})$ 
related to the lattice Euclidean correlators 
$D(\tau,\vec{p})$ as 
\begin{align}
  D(\tau,\vec{p})&=\int d^3xe^{i\vec{p}\cdot\vec{x}} \left\langle J_5(\tau,
  \vec{x})J_5^\dagger(0,\vec{0})\right\rangle 
  = \int_0^\infty {K}(\tau,\omega){A}{(\omega,\vec{p})} d\omega ,
    \label{eq:correlation_function}
\end{align}
where the kernel of the integral transformation $K(\tau,\omega)$ is given by
$K(\tau,\omega)=( e^{-\tau\omega}+e^{-(1/T-\tau)\omega} )/
(1-e^{-\omega/T})$  
with the imaginary time $\tau$.
$J_5(\tau,x)= \bar{c}i\gamma_5 c$ is the local interpolating operator for
the pseudoscalar channel.
We also analyze the dispersion relation of charmonia defined by 
the peak position of the spectral function corresponding to $\eta_c$.

In this work, we perform the MEM analyses in an extended vector space,
which is a product space of two different sets of lattice correlators.
Different lattice correlators measured with the same set of gauge configurations
have a strong correlation among them.
Our analysis takes advantage of the correlation in the MEM analysis.
We find that our method improves the reconstructed images.

\section{\label{sec:mem}Maximum Entropy Method}
Let us first briefly review the basic points of MEM.
To obtain the spectral functions from the lattice Euclidean correlators, 
we have to take 
the inverse transformation of eq.~(\ref{eq:correlation_function}).
MEM reconstructs the most probable image from limited number
of data points for $D(\tau,\vec{p})$ with errors 
on the basis of Bayes' theorem.
In the standard method of least squares, the solution is determined
so as to minimize the value of the chi-square, 
\begin{align}
    \chi^2=&
    \sum_{i,j} \left(D(\tau_i)-D_A(\tau_i) \right)
    C_{ij}^{-1}\left(D(\tau_j)-D_A(\tau_j)\right) ,
    \label{eq:chi_square} 
\end{align}
where the correlations between the lattice correlator with 
different time slices are encoded in the covariance matrix $C_{ij}$.
The discretization of the variables $\tau_i$ is understood. 
However, 
the minimum of $\chi^2$ are degenerating. 
In other words, there exist infinite number of 
spectral images that optimize the $\chi^2$. 
In order to avoid this degeneracy, 
MEM introduces the Shannon-Jaynes entropy \cite{bryan_maximum_1990},
\begin{align}
    S=\int_0^{\infty} \left[ A(\omega)-m(\omega) -A(\omega)\log\left( 
                \frac{A(\omega)}{m(\omega)}
            \right) \right]d\omega ,
    \label{eq:shannon-jaynes_entropy}
\end{align}
where 
the default model $m(\omega)$ expresses the prior knowledge.
Then, we search the most probable image
that maximizes the probability $P(A,\alpha) \sim \exp[Q(A,\alpha)]$ 
with 
\begin{align}
  Q(A,\alpha) = \alpha S(A) - \frac12 \chi^2(A) .
  \label{eq:Q}
\end{align}
The parameter $\alpha$ controls the relative weight between $\chi^2$ and $S$.
The final output image $A_{\rm out}(\omega)$ is then obtained 
by integrating over $\alpha$ and $A$ 
as 
\begin{align}
    A_{\mathrm{out}}(\omega) =\int d\alpha
    \int \left[ dA \right]    {A}(\omega){P(A,\alpha)}.
    \label{eq:Aout}
\end{align}

One of the advantages of MEM is that this method enables us to estimate the error in the reconstructed image
probabilistically. For example, errors can be put on the average of the spectral image $\langle A
\rangle_I$ in some section $I=[\omega_1,\omega_2]$ as 
\begin{align}
    \langle (\delta {A}_{\mathrm{out}})^2 \rangle_I 
    =&
    \int d\alpha 
    \int \left[ dA \right]\int_{I\times I}d\omega d\omega' 
    \delta 
    {A}(\omega)
    \delta {A}(\omega')P(A,\alpha) 
    /  \int_{I\times I}
    d\omega d\omega',
    \label{eq:variance_A}
\end{align}
where
$\delta {A}(\omega)={A}(\omega)-{A}_\alpha(\omega)$ and $A_\alpha$ is the
image which maximizes the probability $P(A,\alpha)$ with some $\alpha$.
The reduction of error is highly desirable to extract physics by MEM analyses.

\section{\label{sec:extension}An extension of MEM}

To analyze observables on the lattice, 
we first generate gauge configurations with Monte Carlo method,
and then perform the measurement on each configuration.
Since the Euclidean correlators with different $\tau$ values on each 
configuration are correlated, the lattice correlators usually have 
strong mutual correlations between different time slices.
In the same way, different correlators, such as those for different
channels or different momenta,  measured on a same set of gauge 
configurations can have strong correlations.

Our strategy is to utilize this correlation in MEM analysis.
For this purpose, we perform the MEM analysis for two or more 
correlators together,
by treating these correlators on the lattice as one vector.
For simplicity, we limit the discussion to the case of 
two correlators, $A_1(\omega)$ and $A_2(\omega)$, in the following.
Equation~(\ref{eq:Q}) is then extended as,
\begin{align}
  Q(A_1,A_2;\alpha_1,\alpha_2)
  = \alpha_1 S(A_1) + \alpha_2 S(A_2) - \frac12 \chi^2 (A_1,A_2) ,
  \label{eq:naraberu}
\end{align}
where $\chi^2 (A_1,A_2)$ is the chi-square including 
the correlation between $A_1$ and $A_2$; 
the correlation between different channels can be 
incorporated by defining the covariance matrix in a usual manner.


In this extended analysis, correlations between the spectral 
images of different channels are incorporated.
When one investigates the difference of two spectral functions,
for example, this method will drastically reduce the error.
It is also expected that the quality of the reconstructed
image in a single channel can be improved, because
information which is not taken into account in the conventional
analysis is newly included in the extended one.
In sec.~\ref{sec:num}, we will see that this is indeed the 
case at least for some cases.

When we perform the extended analysis with eq.~(\ref{eq:naraberu}),
$\alpha_1$ and $\alpha_2$ should be treated as independent 
parameters. The integrals in eqs.~(\ref{eq:Aout}) 
and (\ref{eq:variance_A}) should be extended to two-dimensional ones 
with regard to $\alpha_1$ and $\alpha_2$.
When the correlators for two channels are uncorrelated, 
the extended analysis reduces to two independent MEM analyses
for these channels.
In this exploratory analysis, however, we limit our analysis 
to the case $\alpha_1=\alpha_2$ to reduce the numerical cost,
and take the integrals in this one-dimensional space.
This treatment could lead to incorrect conclusions, because 
$\alpha_\mathrm{max}$'s which maximize the probability 
is different for each correlator when each correlator is separately 
analyzed.
If $\alpha_\mathrm{max}$'s are much different for $\alpha_1$ and $\alpha_2$, 
the reconstructed image with a common $\alpha$ becomes much 
distorted.
An analysis with multi-dimensional $alpha$ is under investigation.

When we perform the extended MEM analysis with several correlators,
however, requirements for numerical analysis become more severe.
First, as the search space becomes larger
the numerical cost for the minimum search becomes higher.
Second, the eigenvalue spectrum of
the covariance matrix is known to show a pathological behavior when
$N_{\mathrm{conf}}$ is not large enough compared to the matrix dimension
$N$ \cite{asakawa_maximum_2001}.
Thus, to perform a reliable analysis, we need more 
configurations as the number of channels increases.
There is another numerical problem.
If the correlation between the correlators is too strong,
the covariance matrix becomes almost singular.
In this case, we need numerically high precision to carry out 
the inversion of the covariance matrix.

\section{Spectral function}
\label{sec:num}

\begin{table}[t]
    \centering
    \caption{\label{tab:lattice_setup}
Lattice simulation parameter}
    \begin{tabular}{llllllll}
        $N_\tau$ & $T/T_{\mathrm{c} }$ & $N_\sigma$ & $L_\sigma[\mathrm{fm}]$ 
        & $a_\tau[\mathrm{fm}]$ & $a_\sigma/a_\tau$ & $\beta$ & $N_{\mathrm{conf}}$\\ \hline
        42 & 1.78 & 64 & 2.496 & 0.00975 & 4 & 7.0 & 427\\
        44 & 1.70 & 64 & 2.496 & 0.00975 & 4 & 7.0 & 407\\
           46 & 1.62 & 64 & 2.496 & 0.00975 & 4 & 7.0 & 401\\
         96 & 0.78 & 64 & 2.496 & 0.00975 & 4 & 7.0 & 207\\
    \end{tabular}
\end{table}

We have analyzed the charmonium correlation functions 
in the pseudoscalar channel
in quenched QCD with the standard Wilson fermion on
an anisotropic lattice with the anisotropy being $4$
\cite{nonaka_charmonium_2011}.
The simulation parameters are summarized in 
Table \ref{tab:lattice_setup} \cite{nonaka_charmonium_2011}.
All results in what follows are obtained with 
the default model $m(\omega)=m_0$ $\omega^2$ with $m_0=1.15$ 
\cite{asakawa_maximum_2001}.

%


\begin{figure}[t]
    \centering
    \includegraphics[width=0.49\textwidth]{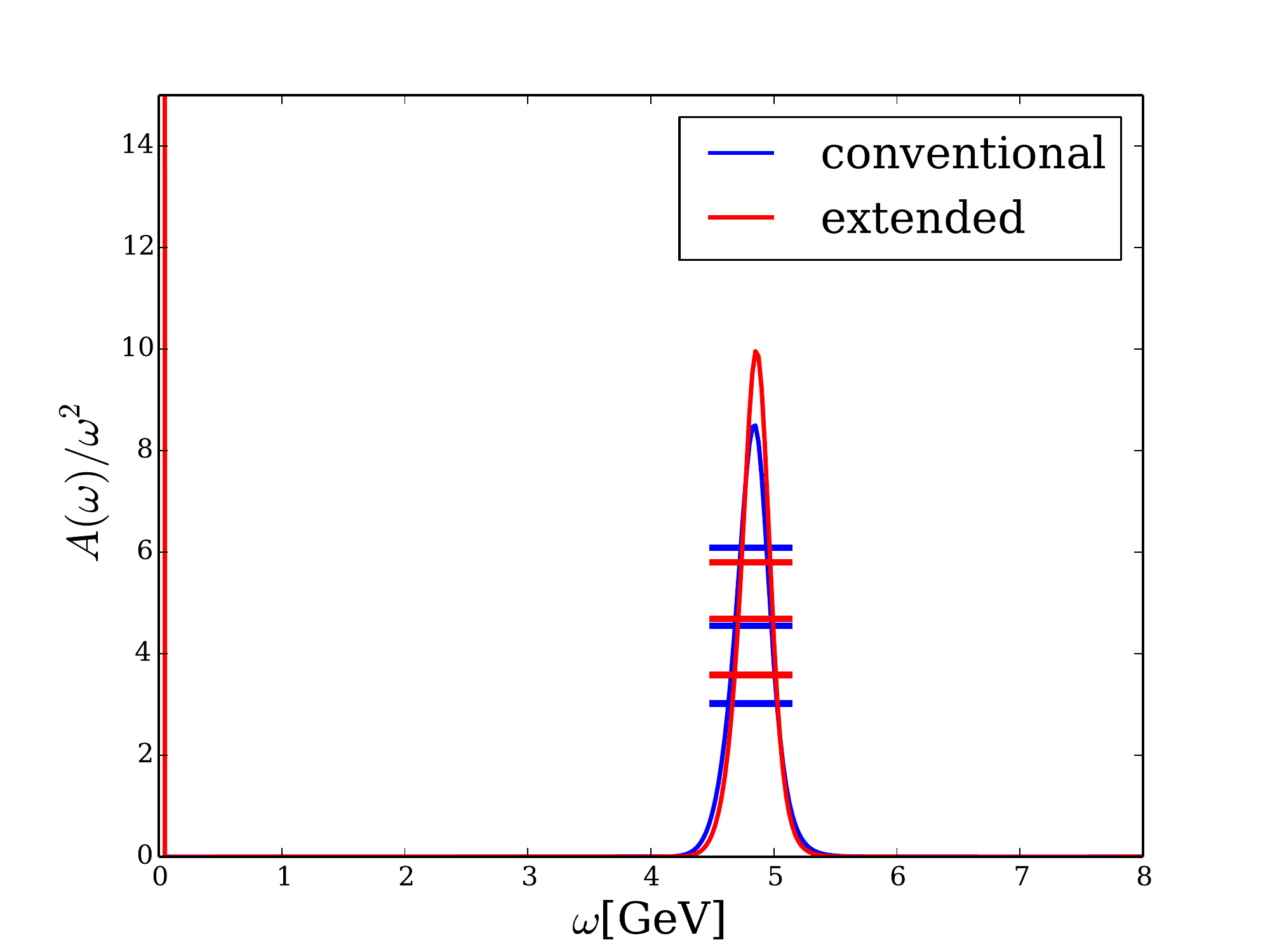}
    \includegraphics[width=0.49\textwidth]{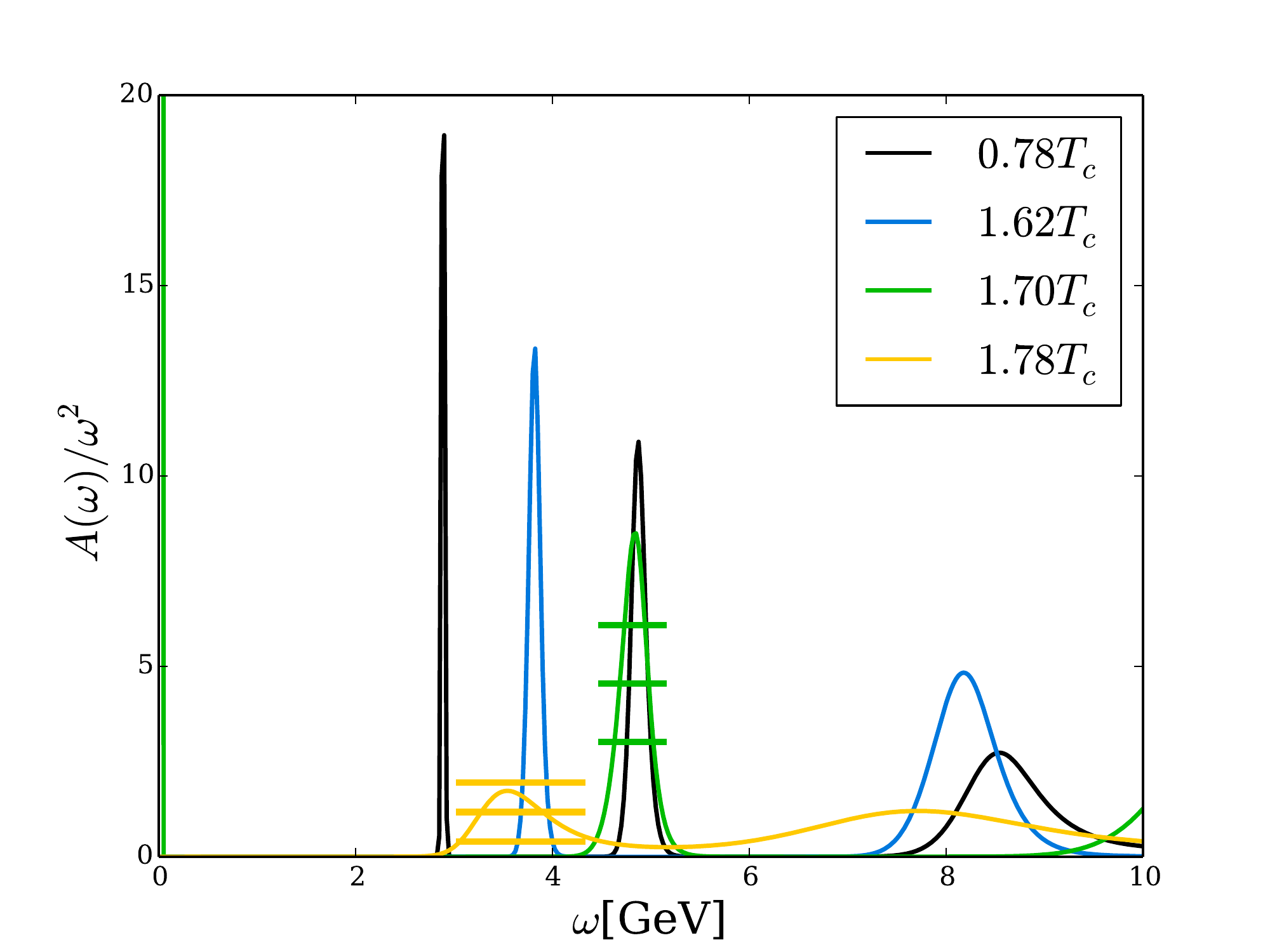}
\caption{
\label{fig:spc}
{\bf Left:} 
Spectral functions for the lowest charmonium 
in the pseudoscalar channel at $T=1.70T_c$ obtained by 
the conventional and extended MEM analyses.
The horizontal lines show the mean value and its errors estimated
by MEM.
{\bf Right:}
Spectral functions in the pseudoscalar channel 
for several values of $T$.
The horizontal lines associated to $T=1.70T_c$ and $1.78T_c$ results 
represent the errors of the peaks.
The survival of $\eta_c$ at $T=1.7T_c$ is indicated.
}
\end{figure}

The left panel of fig.~\ref{fig:spc} shows the result 
for the spectral 
functions with $p=0$ at $1.70T_c$ with the conventional and 
extended analyses.
In the extended analysis, we have used two correlators with 
$p=0$ and $1.5\mathrm~{GeV}$.
The figure shows that the MEM error of the reconstructed image 
obtained by the extended analysis is 
reduced by about $30\%$ and the width of peak becomes narrower
than the one with the conventional one.
This result indicates that the extended analysis is effective 
to improve the quality of the MEM image even when a single 
channel is concerned.
We, however, have found that the error becomes larger 
in another channel at $p=1.5\mathrm~{GeV}$ in the extended analysis.
Our numerical analyses suggest that 
one reconstructed image tends to be improved
while the other tends to be distorted.
This may be caused by the imbalance between the ratio of correlators 
and the ratio of the eigenvectors of the covariance matrices for different 
momenta even in the same channel.
If these ratios are not balanced, one image can be worsen.
We intend to investigate in what situation both images are improved 
and the work in this direction is in progress.

The right panel of fig.~\ref{fig:spc} shows the spectral functions 
at zero momentum for several values of $T$ obtained by the 
conventional method.
The peak corresponding to $\eta_c$ exists for $T\le1.70T_c$,
while the peak disappears at $T=1.78T_c$.
This result suggests that $\eta_c$ still survives at $T=1.70T_c$.

\section{Dispersion relation of $\eta_c$ in medium}

Next,
we investigate the dispersion relation, i.e. $p$ dependence of 
the peak position, of the first peak corresponding to $\eta_c$ 
for $T\le1.70T_c$.
It is, however, not trivial how to define the location of 
the peak and its error in MEM.
For example, the position and width of a peak in a 
spectral image obtained by MEM do not have physical meaning 
and we cannot estimate their errors \cite{asakawa_maximum_2001}.
In the present study, we thus consider the center of the weight of 
a peak
\begin{align}
  {\left\langle \omega\frac{A(\omega)}{\omega^2} \right\rangle_I}/
  {\left\langle \frac{A(\omega)}{\omega^2} \right\rangle_I},
  \label{eq:position}
\end{align}
for an energy interval $I=[\omega_{\rm min},\omega_{\rm max}]$ 
including the peak structure.
Then one can estimate the error for this quantity as 
$ \sigma = \sqrt{\langle \{\omega \delta
        (A(\omega)/{\omega^2})\}^2\rangle_I } /
            \langle A(\omega) / \omega^2 \rangle_I
$
in MEM in a standard way as in eq.~(\ref{eq:variance_A}).

\begin{figure}
    \centering
    \includegraphics[width=9.5cm]{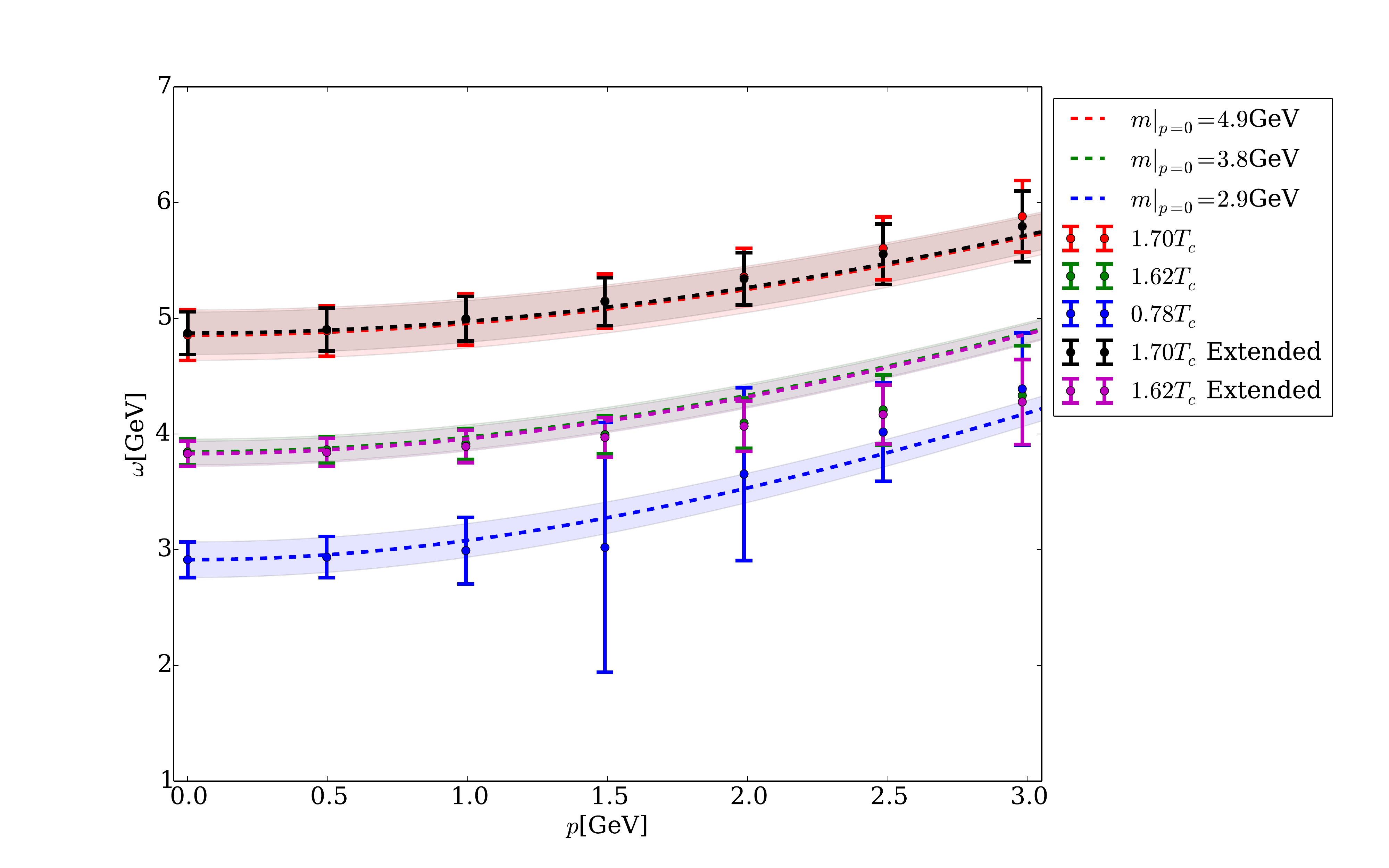}
    \caption{\label{fig:dispersion} 
    Dispersion relation of $\eta_c$ at $T=0.78T_c$, $1.62T_c$ and $1.70T_c$.
    For $T=1.62$ and $1.70 T_c$, we analyze the spectral functions with the conventional MEM
    and extended one.
The dashed lines show the dispersion relation with the Lorentz invariant form
    eq. (\protect\ref{eq:dispersion_Lorentz}).
}
\end{figure}

Figure~\ref{fig:dispersion} shows the result of dispersion relation 
determined by eq.~(\ref{eq:position}) 
with the conventional and extended analyses.
We have checked that the result hardly changes
with the variations of $\omega_{\rm min}$ and $\omega_{\rm max}$.
From the figure, one sees that the peak position at $p=0$
moves to larger $\omega$ as $T$ increases. 
This result suggests that the rest mass of $\eta_c$ 
becomes heavier for higher $T$.
Because of the Lorentz symmetry, 
the dispersion relation in the vacuum takes the form,
\begin{align}
    \omega = \sqrt{m^2 + p^2},
    \label{eq:dispersion_Lorentz}
\end{align}
where $m$ is the mass.
The figure shows that the dispersion relation for $T=0.78T_c$
is consistent with this behavior.
In medium, on the other hand, there is no reason that the dispersion 
relation is given by eq.~(\ref{eq:dispersion_Lorentz}).
Figure~\ref{fig:dispersion}, however, shows that the dispersion 
relation traces eq.~(\ref{eq:dispersion_Lorentz}) well
even at $T=1.70T_c$ within the error, 
which is a highly nontrivial result.

We performed the above analysis both in the the conventional and 
extended methods. Both results are presented in 
fig.~\ref{fig:dispersion} for $T/T_c=1.62$ and $1.70$.
The result shows that the improvement for each error 
is not large for these analyses.
When the correlations between the values at different momenta
are in question, the benefit of the extended analysis will become
prominent.

\section{Discussions and conclusion}

In this study, we analyzed the charmonium spectral function 
in the pseudoscalar channel at nonzero momentum and dispersion 
relations of $\eta_c$ in an extended MEM analysis which utilizes 
the correlation between correlators with different momenta.
A method to define the peak position in a spectral function 
with the error for it in MEM is proposed.
We found that the dispersion relation of the charmonium still follows the 
Lorentz invariant form above $T_c$ until its dissociation.
We point out that this result is not trivial at all.

Analyzing the correlators 
in the extended method,
we found that
our method reduces the error of reconstructed images with MEM.
However, this reduction is limited. 
There is several possible reasons for this.
First, when we adopt the extended covariance matrix to define $\chi^2$, 
the fitting error does not necessarily reduce. 
This does not mean that this analysis is wrong. 
Second, when we extend the correlator space, we need more 
configurations to get a reliable covariance matrix.

In this work, we analyzed only two correlators with two different momenta together.
However, we can use more than three correlators not only with different momenta
but also in ccdifferent channels, for example, pseudoscalar and vector channels, vector
channel with transverse and longitudinal components, and so forth.

\section*{Acknowledgments}
This work was supported in part by 
JSPS KAKENHI Grant Numbers 23540307, 25800148, and 26400272.
The numerical calculations have been performed on the PACS-CS computer at
University of Tsukuba, Blue Gene at KEK and $\phi$ at KMI at Nagoya University.



\end{document}